\documentclass[conference]{IEEEtran}
\IEEEoverridecommandlockouts
\usepackage[T1]{fontenc}
\usepackage{cite}
\usepackage{amsmath,amssymb,amsfonts}
\usepackage{algorithmic}
\usepackage{graphicx}
\usepackage{textcomp}
\usepackage{subcaption}
\usepackage{float}
\usepackage{xcolor}
\usepackage{threeparttable}
\usepackage[normalem]{ulem}
\graphicspath{ {./assets/} }
\def\BibTeX{{\rm B\kern-.05em{\sc i\kern-.025em b}\kern-.08em
    T\kern-.1667em\lower.7ex\hbox{E}\kern-.125emX}}
    
\newcommand{\pkg}[1]{\texttt{#1}}

\begin{document}

\title{Multi-Output Random Forest Regression to Emulate the Earliest Stages of Planet Formation}

\author{\IEEEauthorblockN{Kevin Hoffman}
\IEEEauthorblockA{\textit{University of Virginia}\\
Charlottesville, United States \\
keh4nb@virginia.edu}
\and
\IEEEauthorblockN{Jae Yoon Sung}
\IEEEauthorblockA{\textit{University of Virginia}\\
Charlottesville, United States \\
js2yp@virginia.edu}
\and
\IEEEauthorblockN{André Zazzera}
\IEEEauthorblockA{\textit{University of Virginia}\\
Charlottesville, United States \\
alz9cb@virginia.edu}
}

\maketitle

\begin{abstract}
In the current paradigm of planet formation research, it is believed that the first step to forming massive bodies (such as asteroids and planets) requires that small interstellar dust grains floating through space collide with each other and grow to larger sizes. The initial formation of these pebbles is governed by an integro-differential equation known as the Smoluchowski coagulation equation \cite{Smoluchowski1916}, to which analytical solutions are intractable for all but the simplest possible scenarios. While brute-force methods of approximation have been developed, they are computationally costly, currently making it infeasible to simulate this process including other physical processes relevant to planet formation, and across the very large range of scales on which it occurs. In this paper, we take a machine learning approach to designing a system for a much faster approximation. We develop a multi-output random forest regression model trained on brute-force simulation data to approximate distributions of dust particle sizes in protoplanetary disks at different points in time. The performance of our random forest model is measured against the existing brute-force models, which are the standard for realistic simulations. Results indicate that the random forest model can generate highly accurate predictions relative to the brute-force simulation results, with an $R^{2}$ of 0.97, and do so significantly faster than brute-force methods.
\end{abstract}


\section{Introduction}
\label{sec:intro}
The question of how planets come to be is one which has puzzled the astronomical community for generations, and one which remains partially unsolved. When a dense core of gas and dust collapses under gravity and forms a star, a fraction of the rotating core material forms a protoplanetary disk around the new star. Dust in the disk then settles and interacts with itself and the surrounding disk gas. The coagulation of dust particles from very small (\textless 1 \textmu m) into larger particles ($\sim$1 cm), which subsequently coalesce into larger, gravitating bodies (\textgreater 1 km), is recognized as the first steps towards planet formation. Modeling the time-dependent distributions of dust particle sizes is, however, computationally expensive due to the dynamic nature of the evolution problem. Algorithms to perform this modeling exist \cite{Brauer_dust_2008}, but are prohibitively costly for reasonable use in large-scale dynamical simulations, even if shortcuts and concessions are made \cite{Drazkowska_dynamicsdust_2019}.

In this paper, we develop a random forest regression model which outpaces a brute-force numerical simulation in predicting future size distributions of the dust, while preserving precision. The model predicts the size distribution of dust particles in a protoplanetary disk after a given time interval from input parameters such as the relative velocity between particles, temperature of the gas, total dust density, and the dust to gas ratio.

\section{Related Work}
\label{sec:related}

\subsection{Dust Evolution}
\label{sec:dust-evo}
In the current understanding of planet formation processes, it is the collision and growth of dust particles that seeds the formation of more massive bodies that eventually become planets. There are, however, challenges to getting dust to collide and grow to large sizes. Testi et al. \cite{testi_dust_2014} and Johansen et al. \cite{Johansen_et_al_2014} describe the current understanding of dust collisions in protoplanetary disks based on experimental data, theory, simulations, and observations. One particularly successful and well-used approach to modelling dust evolution was developed by Birnstiel, Dullemond, and Brauer \cite{birnstiel_gas-_2010}, who built upon previous models of size and growth of dust in protoplanetary disks to enable the simulation of the dust evolution over millions of years, albeit with a simplified one-dimensional model. This is the model upon which we've based our research.

\subsection{Machine Learning Approaches for Planet Formation}
\label{sec:ml-approaches}
Researchers have applied machine learning techniques in a few planet formation studies, successfully demonstrating the power of machine learning in similar problems. For example, Auddy and Lin \cite{auddy_machine_2020} predict masses of planets from gaps in protoplanetary disks using a machine learning model. A deep neural network was trained on hydrodynamic simulations with a variety of input conditions, which is then able to flexibly take in a range of input parameters and make accurate predictions. In addition, Alibert and Venturini \cite{alibert_using_2019} developed a deep learning model for predicting the mass of planetary envelopes and the critical mass when these planets accrete gas too quickly. The model was trained on data that was computed with several differential equations. The deep neural networks make accurate predictions with less computing time, and indeed provide more accurate results than available analytical fits.  

\subsection{Multi-Output Random Forest Regression}
\label{sec:multi-rf}
Breiman \cite{Breiman2001} formalized random forest decision trees for classification and regression in 2001, after putting forward a variety of ensemble techniques for decision trees in the 1990s. Random forest regression was then extended to multi-target settings by Kocev, Vens, Struyf, and D\v{z}eroski \cite{kocev_tree_2013}, and has been applied to real problems since. For example, González, Mira, and Ojeda \cite{gonzalez_applying_2016} applied multi-output random forest to predict electricity prices jointly with demand in Spain. The study demonstrated that multivariate response random forests are more accurate than univariate response random forests.

\subsection{Other Implementations}
\label{sec:mdn}
Early in our project, we were interested in trying to leverage neural network architectures to approach this problem. In particular, we felt that the concept of mixture density networks (MDNs), which have shown promise in problems involving multi-target outputs such as ours, could be effective \cite{mdns}. After some time spent developing an MDN architecture for this problem, however, our predictions were insufficiently accurate, and we chose to implement the random forest regression instead\footnote{Code can be made available upon request.}.

We also investigated using gradient boosted trees as our decision-tree based approach, but were unable to find a pre-existing implementation for multi-target gradient boosted trees, and dedicating resources to investigating the development of one was beyond the scope of the project.

\section{Data}
\label{sec:data}

\subsection{Data Generation}
\label{sec:generation}

The data for our model are generated from 10,000 dust growth simulation models in protoplanetary disks across a range of initial conditions. The code for these brute-force models \cite{coag_code} is based on Birnstiel, Dullemond, and Brauer \cite{birnstiel_gas-_2010}. The input parameters for these models include temperature of the gas ($T_{gas}$), the dust to gas ratio ($d2g$), the distance of the disk to the star ($R$), the stellar mass ($M_{star}$), a dimensionless parameter characterizing the level of turbulence ($\alpha$), and the gas surface density ($\Sigma$) \cite{ShakuraSunyaev_alpha_1973}. The surface density and temperature of the gas are calculated using the distance to the star in coordination with simple power laws \cite{Hayashi_solarsystem_1985,Alexander_temperature_2004}. The density, temperature, and level of turbulence control the relative velocity of the dust particles, which determines whether particles will stick together or be destroyed during a collision. The parameter space for these simulations is summarised in Table \ref{tab:param_space}.

\begin{table}[htbp]
\caption{Parameter Space}
\begin{center}
\begin{threeparttable}[b]
\begin{tabular}{|c|c|c|c|}
\hline
 \textbf{Parameter} & \textbf{Range} & \textbf{Units} & \textbf{Count\tnote{a}} \cr 
 \hline
 $d2g$ & $ [1 \times 10^{-4}, 1]$ & -- & $5$ \cr
  \hline
 $R$ & $ [3.16 \times 10^{-1}, 5 \times 10^{2}]$ & $AU$ & $400$ \cr
  \hline
 $M_{star}$ & $1$ & $M_\odot$ & 1 \cr
  \hline
 $\alpha$ & $[1 \times 10^{-5}, 1 \times 10^{-1}]$ & -- & $5$ \cr
  \hline
 $T_{gas}$ & $[4.547 \times 10^1, 1.778 \times 10^3]$ & $K$ & $400$\tnote{b}  \cr
  \hline
 $\Sigma$ & $[1.52 \times 10^{-1}, 9.960 \times 10^3]$ & $g/cm^2$ & $400$\tnote{b}  \cr 
  \hline
\end{tabular}
\begin{tablenotes}
  \item[a] Number of unique values of each parameter.
  \item[b] Derived from $R$. See text for details.
\end{tablenotes}
\end{threeparttable}
\end{center}
\label{tab:param_space}
\end{table}

\begin{figure}[tp]
\includegraphics[width=\columnwidth]{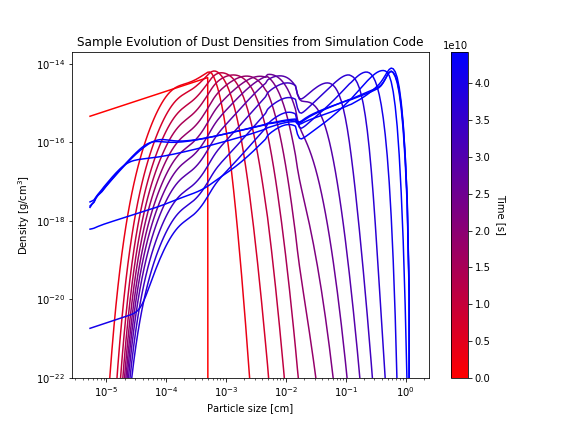}
\caption{Example output of a dust growth simulation over time with multiple snapshots. The initial distribution is shown in red and is the furthest to the left. As the simulation runs through time, the particles mostly clump together and form larger particles evidenced by the size distribution shifting to the right with time. At late times, the particles reach a fragmentation ``size barrier" and start to be destroyed rather than growing any larger.}
\label{fig:dust_sim}
\end{figure}

For a given set of initial parameters, the simulations produce a set of dust densities ($g/cm^3$) in 171 particle size bins ($cm$) for different times (snapshots), up to a maximum of $10^6$ years of evolution, or until the dust distribution reaches a steady state in time. The number of snapshots within a model can vary from a few to thousands, and the difference in time between snapshots can vary from tens of minutes to thousands of years. In Fig. \ref{fig:dust_sim}, we present an example of one model run of the simulation code, and the changing distribution of dust densities over time.

\subsection{Training Data}
\label{sec:train_data}
The training and test data for our model are constructed from the output of the simulation runs. In addition to being able to predict the final dust distribution from a simulation run, we also want our model to be able to predict between any two points in time. For each model, we sampled snapshots with replacement, to get up to 190 distinct pairs of snapshots. The first and last snapshot are always included for each model. For each pair of snapshots, the time of the first snapshot $t$ (seconds) and the difference between the two snapshots $\Delta_t$ (seconds) is recorded. The dust density distributions for each snapshot are normalized by dividing each bin by the sum of the densities in all bins. After normalization, any normalized value less than $10^{-30}$ is set to 0, as dust densities less than this threshold are not physically relevant. The input features for each observation are the 6 initial condition parameters described earlier, $t$, $\Delta_t$, and the 171 input normalized dust density bins. The second snapshot, or the output snapshot after $\Delta_t$, of the 171 normalized dust density bins is the target for prediction. This process is repeated for every pair of snapshots selected for one model, and repeated for all 10,000 models. This sampling process of the models yields approximately 1.4 million observations. Our data set was uniformly and randomly split into $90\%$ training data and $10\%$ test data.

\section{Random Forest Implementation}
\label{sec:random}
\subsection{Background}
\label{sec:RFtheory}
Random forest regression is a supervised machine learning algorithm, which predicts an outcome from a combined pooling of base estimators in a decision tree. Each tree is built from independently randomly sampled vector values. 

The model uses all input features of the training data and tests all possible combinations of dividing the samples in any leaf nodes into two parts. For each proposed split, it calculates the mean squared error (MSE) between the dividing line and every sample, and selects the split with the lowest MSE. The splitting process continues on all leaf nodes until it reaches the maximum depth for each branch, or further splitting of leaf nodes would reduce the number of samples in the leaf below the minimum threshold. The final prediction is the average of the predictions from each tree \cite{Breiman2001}.

Since our data are very non-linear and multi-dimensional, decision tree regression appeared to be a good choice for our model. A classic concern of employing decision trees, however, is overfitting. Breiman \cite{Breiman2001} proves that, as a consequence of the Law of Large Numbers, the overfitting problem is avoided by the random forest ensemble.

As a result of the setup of the project, the model needed to be able to predict in many dimensions. The \pkg{ensemble} module from Python's \pkg{scikit-learn} library includes multi-output random forest regression as a meta-estimator, which we use \cite{scikit-learn}.

\subsection{Tuning}
\label{sec:tuning}
As referenced in section \ref{sec:intro}, one of the key advantages of the random forest implementation is the relative ease of tuning. We focused our tuning on a triplet of hyperparameters: \pkg{n\_estimators} (the number of trees), \pkg{max\_depth} (the maximum depth of a trees), and \pkg{min\_samples\_leaf} (the minimum number of samples which need to be included for a node to become a leaf). More information on the hyperparameters can be found in the documentation for \pkg{scikit-learn}\footnote{https://scikit-learn.org/stable/modules/generated/ \allowbreak sklearn.ensemble.RandomForestRegressor.html} \cite{scikit-learn}. We chose these hyperparameters for tuning based in part on prior research in \cite{segal-xiao-mrf} which suggests the significance of the maximal depth of the trees, and in part based on our own understanding of how the minimal samples for a leaf node would affect our model size and performance.

For tuning, we used $R^{2}$ (the proportion of change in response variable which can be explained by the model) as a performance metric, enabling us to quickly and easily compare models' predictive capacity. Additionally, we needed to ensure that the fully trained model was of a reasonable size (i.e., one that could fit in memory) for download and use by other researchers, who may not have supercomputer access. As a result, we scored our models in part based on the size of the output file, and sought to balance the trade-off between a regression model which performed at maximal accuracy, and one which could be reasonably made available to other researchers.

We manually experimented with values for hyperparameters, using powers of 2 for \pkg{min\_samples\_leaf}. In the end, we found that 1000 \pkg{n\_estimators}, with a \pkg{max\_depth} of 30, and \pkg{min\_samples\_leaf} of 256\textemdash which yielded a 6.22 GB model with an $R^{2}$ of 0.97\textemdash were sufficiently well-performing hyperparameters \footnote{Our best model performed even better, but was too large to be used (over 30 GB).}.

\section{Results}
\label{sec:result}

\subsection{Performance}
\label{sec:performance}

In order to quantify how well the trained model performs, we used the Jensen-Shannon divergence (JSD), defined as:
\begin{equation}
    \text{JSD}(P\| Q) = \frac{1}{2}\text{D}(P\|M) + \frac{1}{2}\text{D}(Q\|M)
\label{JSD}
\end{equation}
with D being the Kullback-Leibler divergence, and M as one-half of the sum of P and Q \cite{1365067}. Unlike Kullback-Leibler divergence, however, Jensen-Shannon is symmetrized and bounded by 0 and 1.
When the divergence is 0, it indicates that the prediction is identical to the actual output. On the other hand, a score of 1 means that the actual value is maximally different.  

\begin{figure}[bp]
\includegraphics[width=\columnwidth]{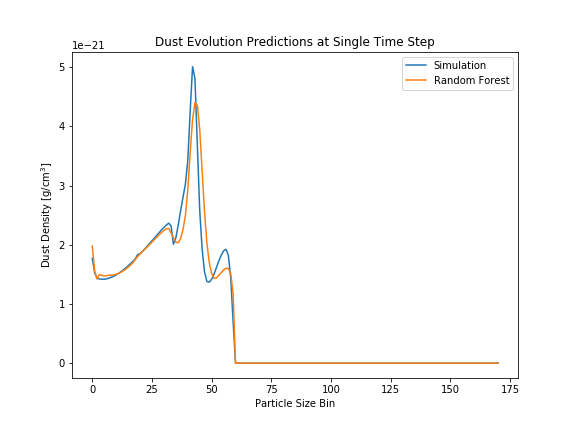}
\caption{A good prediction after a time interval of approximately 377391 years, showing the dust density as a function of bin number. The JSD for this output is about 0.061.}
\label{fig:good prediction}
\end{figure}

One of the most significant features of the data is the location of the distribution peak, and accurately predicting this was an important target for our model. 
After analyzing the JSD scores and the distributions, we defined the predictions to be ``good" if JSD $\leq0.2$, a threshold we determined by visual examination of sample predictions and outputs. Models with this JSD constraint tended to predict the location and height of the peak accurately. 86\% of the predictions from our model are classified as ``good". An example output of a good prediction is provided in Fig. \ref{fig:good prediction}.

\begin{figure}[bp]
    \includegraphics[width=\columnwidth]{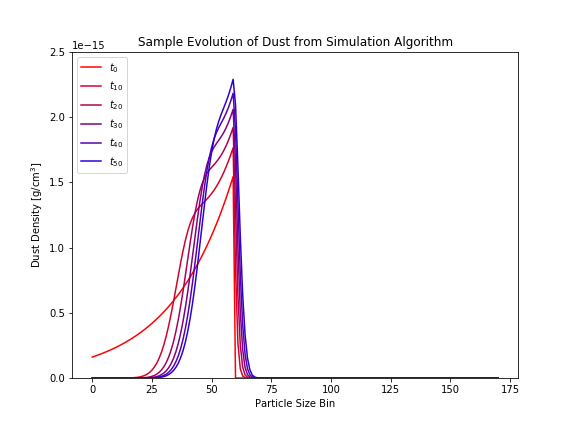}
    \caption{Brute-force simulation outputs of the dust distribution as a function of particle size bins over several time steps (shown with differently colored lines).}
    \label{fig:dust_sim_accum}
\end{figure}

One of the advantages of our implementation is its ability to make a prediction for some arbitrary time in the future by specifying the $\Delta_t$ parameter, while the brute-force method has to simulate the entire evolution track. In Fig. \ref{fig:dust_sim_accum}, we present a representative evolution of dust distributions over time using the brute-force simulation code. 

\begin{figure}
     \begin{subfigure}[htbp]{\columnwidth}
         \includegraphics[width=\columnwidth]{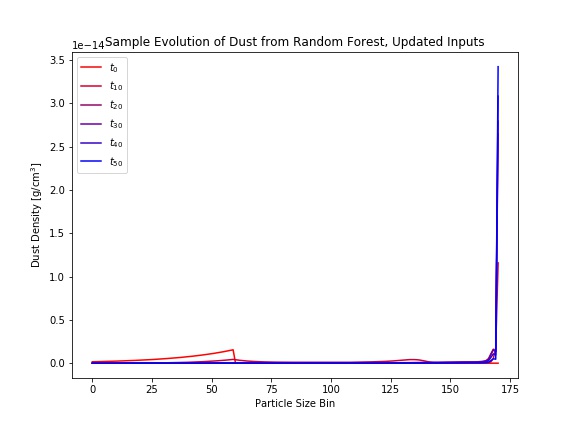}
         \caption{A prediction for the same input parameters as in Fig. \ref{fig:dust_sim_accum}, with the same time steps, but using each prediction output as the next prediction's input.}
         \label{fig:pred_accum}
     \end{subfigure}
     \hfill
     \begin{subfigure}[htbp]{\columnwidth}
         \includegraphics[width=\columnwidth]{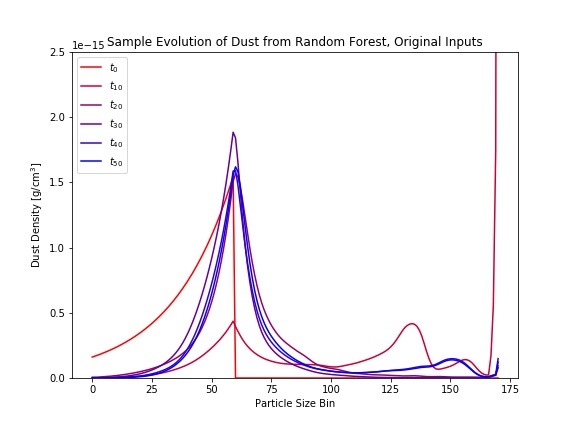}
         \caption{A prediction for the same model, but using the original input parameters and accumulated time steps for the next prediction's input.}
         \label{fig:pred_no_accum}
     \end{subfigure}
        \caption{When the random forest uses the outputs of one time step prediction as the inputs for the next, errors rapidly compound. This can be solved by making predictions from the original input parameters.}
        \label{fig:accum_errors}
\end{figure}

When we try to sequentially predict the same model over several time steps, using the output of one time step prediction as the input for the next prediction (as the simulation code does internally), we see in Fig. \ref{fig:pred_accum} that errors rapidly compound and the distribution quickly accumulates on the right as the random forest predicts rapid coagulation of dust particles. If, instead, we have the random forest predict from the initial conditions for each time interval (i.e. the accumulated time steps), we produce the predictions seen in Fig. \ref{fig:pred_no_accum}. While the predictions trend rightwards for the first time steps, the model then returns to predicting a correctly-located peak thereafter.

The principal motivation for our random forest model is that it needs to make faster predictions than the existing brute-force methods. Being able to make faster predictions for dust evolution would decrease the cost of running large-scale planet formulation simulations that include dust. For the 10,000 model simulations we used, the fastest run time was 0.2 seconds, the median run time was 0.6 seconds, and the slowest run time was 60,207.9 seconds. The median time for our random forest model to make one prediction for an arbitrary time in the future is 0.13 seconds, approximately an 80\% decrease over the median simulation time. Furthermore, the run time of a single simulation depends on the input parameter space, while the time to make a prediction for our random forest model is independent of the input parameters.

\subsection{Model Errors}
\label{sec:error}

\begin{figure}[tp]
\includegraphics[width=\columnwidth]{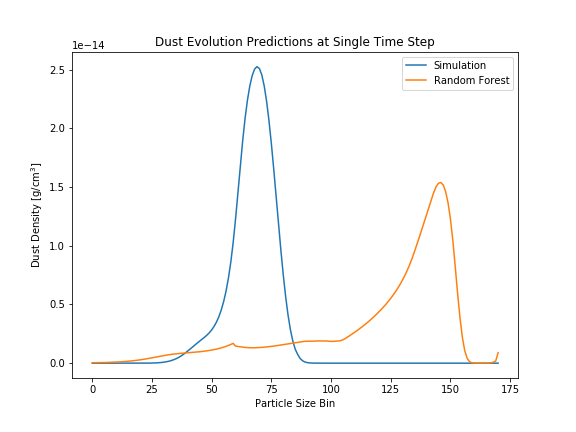}
\caption{Sample output of a bad prediction from the random forest model. We see that the true distribution has its peak below size bin 75. The predicted distribution has its peak much further to the right, indicating that the model predicted much more dust coagulation. The JSD for this output is about 0.712.}
\label{fig:bad_pred}
\end{figure}

Despite the generally high performance of the model's predictive capabilities, there are certain exceptions, where the random forest model we have implemented will make distribution predictions which are significantly different from the ground truth of the simulation output. An example output of one such prediction, made for a single time step, is provided in Fig. \ref{fig:bad_pred}. 

\begin{figure}[bp]
\includegraphics[width=\columnwidth]{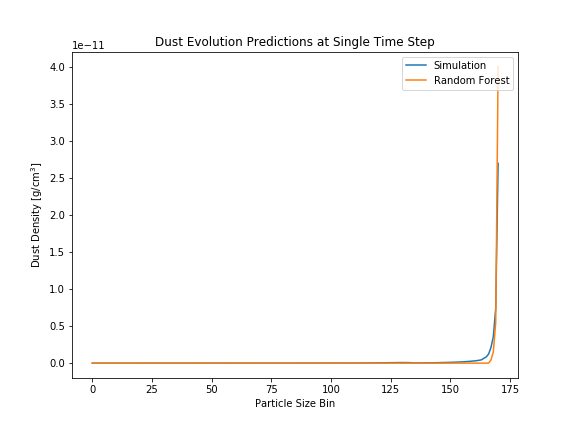}
\caption{A prediction after a time interval of about 607 years, which exhibits heavy accumulation in the high bin numbers (corresponding to large dust sizes). The JSD for this output is around 0.284.}
\label{fig:right_accum}
\end{figure}

During our analysis, we found that many of the outputs with low accuracy predicted too much coagulation relative to the accepted true distribution, thus leading to a peak in the density distributions too far to the right. We suspect this is because a number of the training examples accumulate all of the dust in high-size bins, as in Fig. \ref{fig:right_accum}, which caused the random forest to over-predict such occurrences.

In order to be able to predict whether the algorithm will produce what we consider to be a ``good" or ``bad" prediction, we chose to analyze the relationship between the distance between predicted and actual modes of the distributions (which we call ``bin error") and JSD.

\begin{figure}[tp]
\includegraphics[width=\columnwidth]{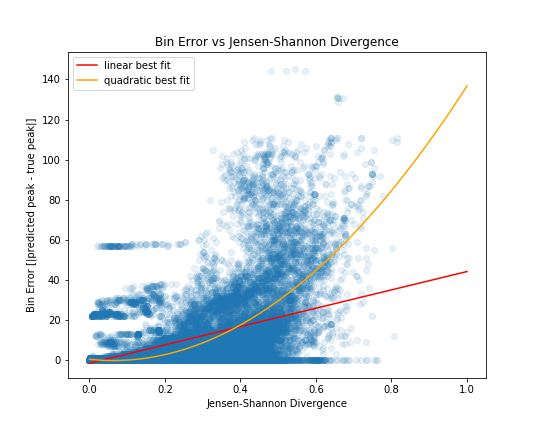}
\caption{Comparing the difference in predicted and true bin number of the density peak of distributions to the reported JSD for that same output (plotted as blue circles), and overlaying both linear and quadratic best fit curves over top.}
\label{fig:bin_error}
\end{figure}

We see in Fig. \ref{fig:bin_error} that there are some models with a high bin error but a low divergence. This occurs in cases where the predicted distribution is the right shape (relative to the expected result), but centered incorrectly. We also observe that both the linear and quadratic fits are being pulled down by a preponderance of points with 0 bin error, but with a spread of divergence values. This indicates models where the predicted distribution has the correct center, but incorrect shape. The quadratic best fit also shows that, as the bin error increases, the rate of change of bin error with respect to divergence grows. This indicates that JSD is a good evaluation metric especially for our \emph{best} predictions (those in the lower left corner of the plot), and gives us more information than merely looking for the distance between peaks, i.e., it tells us if the shape of the predicted distribution might be incorrect.

To address these sub-optimal predictions, we built a gradient boosting classifier using XGBoost \cite{Chen:2016:XST:2939672.2939785}. We trained the additional classifier on 90\% of the previous test set, which we labelled as being ``good" or ``bad" predictions as described in section \ref{sec:performance}. This classifier has an accuracy of 98\%, and will raise a warning to users if the prediction is classified as being of poor quality.

\section{Conclusion}
\label{sec:conclusion}
We have developed a random forest regression model which significantly outperforms currently existing brute-force methods for simulating dust coagulation with respect to computation speed, while maintaining high precision. 86\% of our predictions had a very low Jensen-Shannon divergence relative to the brute-force method, giving us confidence in the efficacy of our implementation.

Some mention has already been made in this paper to areas in which we feel that more exploration is warranted. In particular, further examination of leveraging new developments in neural networks, or using other decision tree regression models, would be a useful extension of this work.

Our existing framework has been limited to the problem of predicting dust coagulation behavior based on static initial conditions, but could possibly be neatly extended to train on and predict over dynamical tracks of input parameters, which would represent more realistic scenarios.

With our method, researchers can now run large-scale dynamical simulations of planet formation at a lower cost. Additionally, dust evolution is of interest to other astronomical researchers, who may be studying such phenomena as dust behavior in interstellar space, molecular clouds \cite{Hirashita_interstellarturbulence_dust_2009MNRAS}, and even in other galaxies \cite{Aoyama_galaxy_dust_2020}, and we believe that our methods can be deployed in related work with minimal adjustments.

Interested researchers can make use of our model via the \pkg{astrodust} Python package\footnote{www.github.com/kehoffman3/astrodust}.

\section{Acknowledgments}
\label{sec:acknowledgments}
The authors wish to thank Jonathan Kropko for his support, encouragement, flexibility, and mentorship throughout the past year; as well as Jonathan Ramsey and Ilse Cleeves for their tireless guidance, wisdom, astronomical knowledge, and enthusiasm for our team's project.
Additionally, we would like to acknowledge Sebastian Stammler, University Observatory, Faculty of Physics, Ludwig-Maximilians-Universit\"at M\"unchen, Germany, for providing the brute-force simulation code which generated our training data.

\bibliographystyle{./bibliography/IEEEtran}
\bibliography{./bibliography/IEEEabrv,./bibliography/IEEEexample,./bibliography/Paper}

\end{document}